\documentclass{Interspeech2024}
\usepackage{subcaption}

\interspeechcameraready

\title{A cost minimization approach to fix 
the vocabulary size in a tokenizer for an End-to-End ASR system} 

\name{Sunil Kumar}{Kopparapu}
\name{Ashish}{Panda}

\address{
  TCS Research, India}
\email{\{sunilkumar.kopparapu,ashish.panda\}tcs.com}


\def\corpus{{\cal S}}
\def\words{{w}}
\def\tokenize{{\cal T}}
\def\ksentence{{k}}
\def\sentenceset{{\corpus}}
\def\sent{{s}}
\def\encode{{enc^{\tokenize}}}
\def\decode{{dec^{\tokenize}}}
\def\Ct{\theta_t}
\def\fC{f}
\def\tokensincorpus{\Gamma}
\def\x{\tau}
\def\Cost{{C}}
\def\C{{\Cost}}
\def\sub-words{{tokens}}
\def\sub-word{{token}}
\def\SP{{SentencePiece}}
\def\unigram{Unigram}
\def\bpe{BPE}

\begin{document}

\maketitle

\begin{abstract}

    Unlike hybrid speech recognition systems where the use of \sub-words\ 
    was restricted to phones, biphones or triphones the choice of \sub-words\ in the end-to-end ASR systems is derived from the text corpus of the training data. The use of tokenization algorithms like Byte Pair Encoding (BPE) and WordPiece is popular in identifying the \sub-words\ that are used in the overall training process of the speech recognition system.  Popular toolkits, like ESPNet use a pre-defined vocabulary size (number of tokens) for these tokenization algorithms, but there is no discussion on how vocabulary size was derived. In this paper, we build a cost function, assuming the tokenization process to be a black-box to enable choosing the number of tokens which might most benefit building an end-to-end ASR. We show through experiments on LibriSpeech 100 hour set that the performance of an end-to-end ASR system improves when the number of \sub-words\ are chosen carefully.
\end{abstract}

\section{Introduction}
\label{sec:introduction}
It was a standard practise to choose mono-phones, bi-phones, tri-phones as the \sub-words\ to train a hybrid automatic speech recognition (ASR) system \cite{raissi2021consistent}. The \sub-words\ to be trained was dependent on the prominent sounds in that language and the training data required the phonetic transcription of the training speech corpus. However, with the advent of end-to-end systems and the availability of phonetic transcripts, the move has been on automatically identifying the \sub-words\ from the text data of the training corpus. As a result, most current ASR systems model \sub-words\ derived from the training text rather than use unit which have relevance to the language or pronunciation.

Several  
tokenization algorithms, such as Byte Pair Encoding (BPE) \cite{sennrich-bpe}, WordPiece \cite{wordpiece}, or unigram language model tokenization \cite{unigram} have been researched.  These algorithms, broadly work on the principle of iteratively merging frequently occurring pairs of characters or \sub-words\ to create a vocabulary of tokens that can represent the language's vocabulary efficiently. The difference among these tokenization algorithms lies in the way characters are paired. BPE uses a pre-tokenizer to split the training data into words and creates a set of base vocabulary consisting of all symbols that occur in the set of unique words. After this, BPE learns the merge rules to create a new symbol from two symbols of the base vocabulary. This process goes on till the desired number of symbols (or tokens) are created. While BPE chooses the most frequent symbol pairs to merge, WordPiece merges the symbol pair that maximizes the likelihood of the training data until the desired number of symbols have been obtained. In contrast to BPE and WordPiece, Unigram language model tokenization starts from a large set of symbols and trims down each symbol to obtain a smaller set of symbols. SentencePiece \cite{kudo2018sentencepiece} is a popular language independent sub-word tokenizer and detokenizer for Neural Text Processing. It does not need a pre-tokenizer unlike BPE and hence it is suitable for languages such as Japanese and Chinese. While the above tokenization algorithms work on training text, various efforts have been made to bring in acoustic perspective into the tokenization process. Pronunciation Assisted Subword Modeling (PASM) \cite{pasm}, Acoustic Data Driven Subword Modeling (ADSM) and Phonetically Induced Subwords \cite{pis} are examples of such efforts. 

 Most ASR systems that use sub-word tokens for training fix the number of tokens (vocabulary size) and use one of the above tokenizers to generate the tokens from the training text data.  To the best of our knowledge there has been no discussion on the criteria used for determining  the optimal number of tokens and they are fixed empirically. In this paper, we explore a formulation that can help identify the number of \sub-words\ best suited for ASR training. We set about the task assuming the availability of a sub-word tokenizer and use it as a black box.

We first formulate a cost function which when minimized results in an optimal  number of sub-word tokens for a given training text data. We specifically use an off-the-shelf tokenizer for the purposes of demonstration but it should be kept in mind that the formulation should work for any tokenizer. We then evaluate the performance of a standard deep architecture ASR to validate the choice of the number of sub-word tokens. The main contribution of this paper is in (a) formulating a framework to identify the number of sub-word tokens and (b) building a cost function to enable identifying the optimal number of sub-word tokens and (c) evaluating the performance on a ASR to validate the need for formal identification of the number of sub-word tokens required to train an ASR system.

The rest of the paper is organized as follows. We describe the formulation of the cost function to allow identification of the optimal number of \sub-words\ given a training text corpus in Section \ref{sec:problem}. In Section \ref{sec:experiments} we experiment with LibriSpeech 100 hours training data set to first identify the optimal number of \sub-words\ and then use the \sub-words\ to evaluate the performance of an end-to-end ASR. We  conclude in Section \ref{sec:conclude}.

\section{Problem Setup}
\label{sec:problem}

Let $\corpus$ be a text corpus consisting of  
$\sentenceset = \{ \sent_1, \sent_2, \cdots, \sent_{\ksentence}\}$ $\ksentence$ sentences associated with training data which consists of $\words$ words ($\words_u$, unique). For simplicity, we will assume the text to be English so that blank spaces represent word boundaries and newlines identify sentences. Let $\tokenize$ be a tokenization routine (for example, byte pair encoding \cite{zouhar-etal-2023-formal}) which takes as input  a variable $n$ and operates on $\corpus$ to produce a set of 
tokens $T_n$, namely, 
\begin{equation}
\tokenize(n,\corpus) = T_n
\label{eq:tokenize}
\end{equation}
where $T_n = \{t_1, t_2, \cdots, t_n\}$, $|T_n| = n$ is the number of tokens, $t_i$ is the $i^{th}$ token,  and $t_i \ne t_j$ for $\forall i \ne j$.
Let $\encode$ and $\decode$ be a pair of functions associated with $\tokenize$  such that $$\encode(\sent_i) = \bigcup_{l=1}^{\beta_i} \tau_l$$ acts on a sentence $\sent_i$ in the corpus $\corpus$ and represents it using $\beta_i$ tokens $\{\tau_l\}_{l=1}^{\beta_i} \in T$, and $\decode$ function 
$$\decode\left (\bigcup_{l=1}^{\beta_i} \tau_l \right ) = \sent_i$$
reconstructs $\sent_i$ by concatenating (represented by $\bigcup$) a sequence of tokens. 
Let 
\[
\Ct = \sum_{i=1}^{\ksentence} \beta_i
\]
denote the number of tokens required to span the corpus $\corpus$.
As mentioned earlier,  $\beta_i$ represents the number of tokens required to represent the sentence $\sent_i$. Let $\tokensincorpus$ represent the set of all token required to span the corpus $\corpus$, namely, 
\[
\tokensincorpus = \bigcup_{i=1}^{k} \left ( \bigcup_{l=1}^{\beta_i} \tau_l \right )
\]
where the total number of tokens in $\tokensincorpus$ is 
$| \tokensincorpus | = \sum_{i=1}^{k} \beta_i.$
We can compute the frequency of occurrence of \sub-word\ 
$\tau \in T_n$ 
as
\[
\fC(\x) = \sum_{i=1}^{|\tokensincorpus|} [\tokensincorpus_i = \x].
\]
where $\x \in T_n$ and $\tokensincorpus_i \in \tokensincorpus$ and 
$[\cdot]$ are the Iverson brackets such that $[Q]$ is defined to be $1$ if $Q$ is true, and $0$ if it is false.
Further let $\fC^+$ and $\fC^-$ represent the average of top five most frequently occurring 
tokens and 
the top five 
most infrequently occurring 
tokens respectively.  Note that (a) $\Ct$, (b) ${\fC^+}$, and (c) ${\fC^-}$  are a function of $n$ as seen in 
(\ref{eq:tokenize}).

We hypothesize that choosing the optimal number of tokens 
would be equivalent to finding an $n^*$ which minimizes the cost function (\ref{eq:mimimize}), namely,
\begin{equation}
   n^* = \min_{n} \left \{ \C \right \} 
   \label{eq:nstar}
\end{equation}
where  
\begin{equation}
     \C =  \left \{ \alpha_1 \overbrace{n}^{t_1} + 
     \alpha_2 \overbrace{\left (\frac{\fC^+}{\fC^-} - 1 \right )}^{t_2} + 
     \alpha_3 \overbrace{\left (  \frac{\Ct}{\words} - 1\right )}^{t_3} \right \}
     \label{eq:mimimize}
\end{equation}
is the cost function and $\alpha_{1,2,3}$ are weights which are chosen heuristically.

The first term, $t_1$, in the cost function (\ref{eq:mimimize}) is to ensure that the total number of \sub-words\ used to represent the corpus $\corpus$ is small because a large $n$ means not only training for a large number of \sub-words\ which in turn requires a larger amount of training data but smaller number of \sub-words\ can result in faster training and inference times for ASR systems. With fewer \sub-words\ to model, the computational complexity of the ASR system can be reduced, leading to quicker training convergence and real-time performance during inference.
The second term, $t_2$ ensures  a balance between the most frequently occurring and the least frequently occurring \sub-words, because an imbalance in data can lead to bias during training \cite{katare2024analyzing}. Note that a perfectly balanced training data would have  $\left ( \frac{\fC^+}{\fC^-} \right ) = 1$.
And the third terms, $t_3$ ensures that the number of tokens required to represent the corpus $\corpus$ is close to the number of words $\words$ in the corpus. This constraint makes sure that the cost of compute is minimum, because most major Gen AI portals charge for their services based on the number of tokens required to represent the input and generate an output. 

The construction of the cost function as mentioned in (\ref{eq:mimimize}) to identify the optimal number of \sub-words\ is one of the main contributions of this paper. 

\section{Experimental Setup}
\label{sec:experiments}
We use the LibriSpeech-100\footnote{https://www.openslr.org/resources/12/train-clean-100.tar.gz} (sentences:28537; words:990093)
dataset in our experiments. Note that the database consists of 100 hours of read English speech accompanied by the textual transcript. The training data consists of $\ksentence=28538$ sentences and $\words=990093$ words of which $\words_u= 33798$ are unique. Our experimental evaluations, in this paper, are of two types. We make use of the text transcript and determine the optimal number of tokens. We also validate the usefulness of the obtained optimal number of tokens  by ASR systems using the same tokens. We report the performance obtained on the test-clean\footnote{https://www.openslr.org/resources/12/test-clean.tar.gz} (sentences:2620; words: 52576)
    and test-other\footnote{https://www.openslr.org/resources/12/test-other.tar.gz} (sentences:2939; words:52343) datasets. 

We have used state-of-the-art conformer encoder-decoder architecture as the ASR. The conformer model is 
implemented using 
an ESPNet toolkit \cite{watanabe18_interspeech} with Librispeech 100 hours (low-resource) recipe. 
While we changed the number of \sub-words\ in the original recipe, we used the rest if the model hyper-parameters as described below: 

The encoder of the conformer has $12$ layers and the decoder has $6$ layers. The model dimension is $256$ while number of attention heads used is $4$. Models were trained using Adam optimizer \cite{kingma2017adam} with $\beta_1 = 0.9$, $\beta_2 = 0.98$ and $\epsilon = 10^{-9}$, which is along the lines of the optimizer proposed in \cite{vas_attn}. Warm up steps used is $25000$. The models are trained for $100$ epochs and the batch size was $64$. A single Nvidia RTX 3090 GPU was used.

No language model has been used for shallow fusion in this work. The $n$ 
tokens extracted from the training text served as output units. Features used for the model are log mel spectrograms with $80$ dimensions along with the pitch (total $81$). 
$3$ way speed perturbation \cite{ko15_interspeech} with speed factors of $0.9$, $1.0$ and $1.1$ and SpecAugment \cite{park19e_interspeech} was used in all the experiments.

\section{Experimental Results}
\label{sec:results}

\begin{figure*}[!htb]
    \centering
    \begin{subfigure}{0.475\textwidth}
    \includegraphics[width=\textwidth]{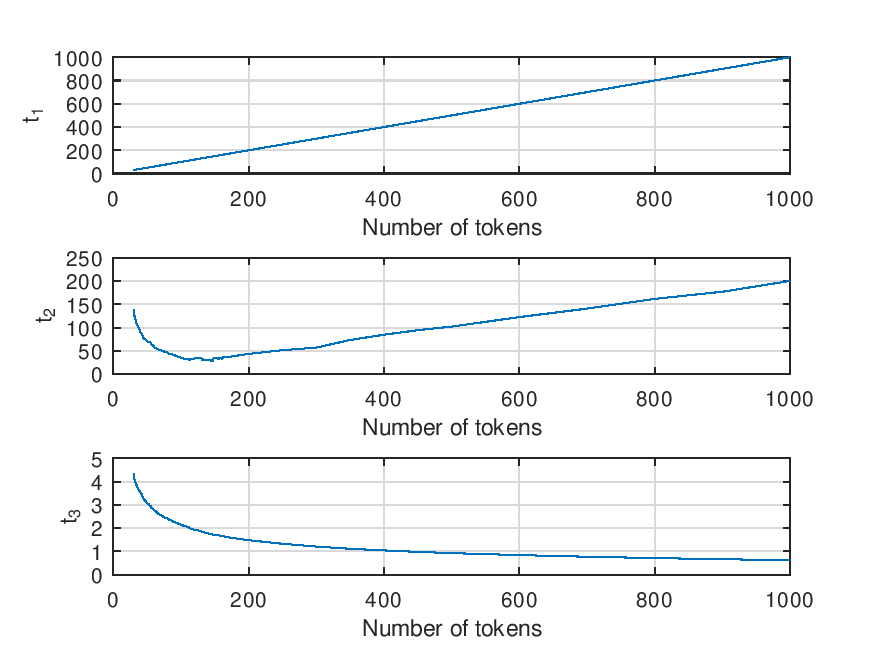}
    \caption{\SP-\unigram.}
    \label{fig:all_data_unigram}
\end{subfigure}
\hfill
        \begin{subfigure}{0.475\textwidth}
    \includegraphics[width=\textwidth]{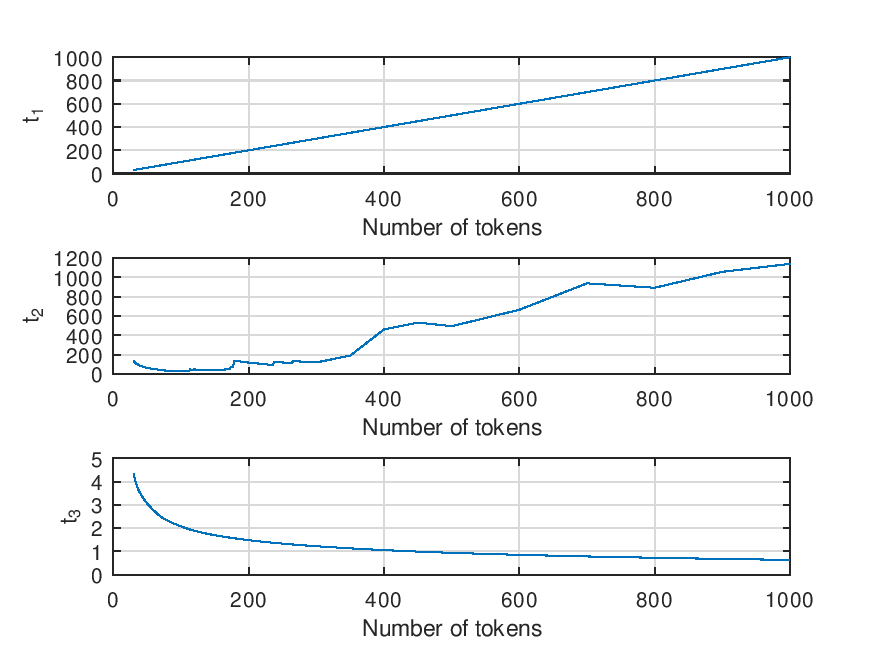}
    \caption{\SP-\bpe.}
    \label{fig:all_data_bpe}
\end{subfigure}
    \caption{$t_1, t_2, t_3$ in the cost function $C$ (\ref{eq:mimimize}) for $n=30$ to $1000$.}
    \label{fig:all_data}
\end{figure*}

\begin{figure*}[!htb]
    \centering
    \begin{subfigure}{0.33\textwidth}
    \includegraphics[width=\textwidth]{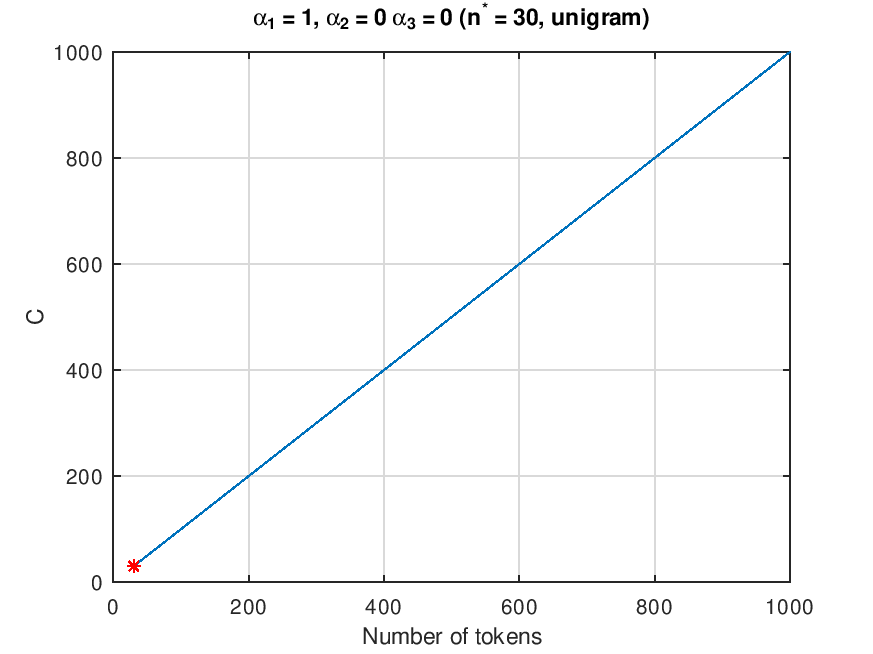}
    \caption{ $\alpha_{1,2,3} = (1,0,0)$; $n^*=30$.}
    \label{fig:nstarsingle_unigram_a}
    \end{subfigure}
    \hfill
    \begin{subfigure}{0.33\textwidth}
    \includegraphics[width=\textwidth]{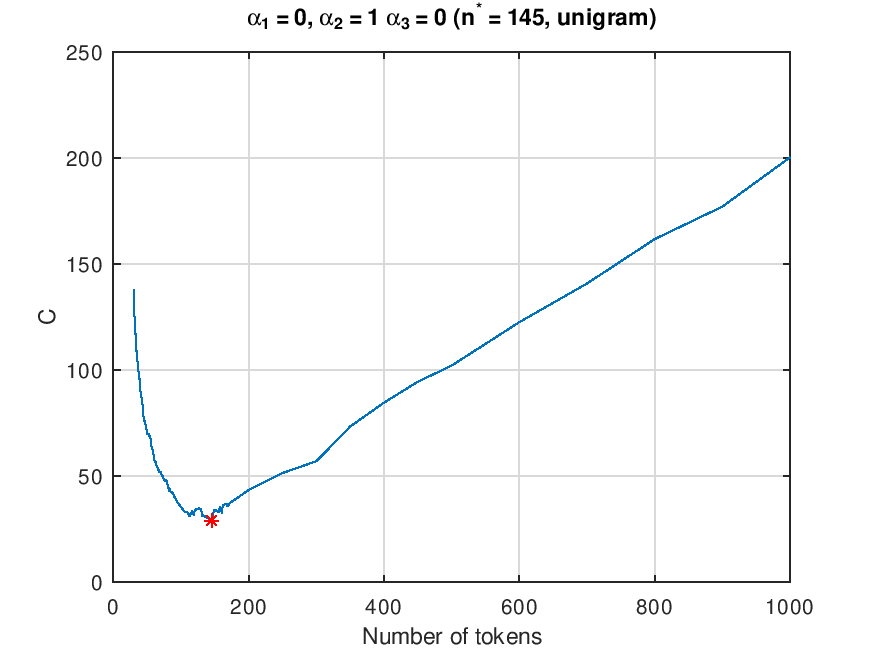}
    \caption{ $\alpha_{1,2,3} = (0,1,0)$; $n^*=145$.}
    \label{fig:nstarsingle_unigram_b}
    \end{subfigure}
    \hfill
    \begin{subfigure}{0.33\textwidth}
    \includegraphics[width=\textwidth]{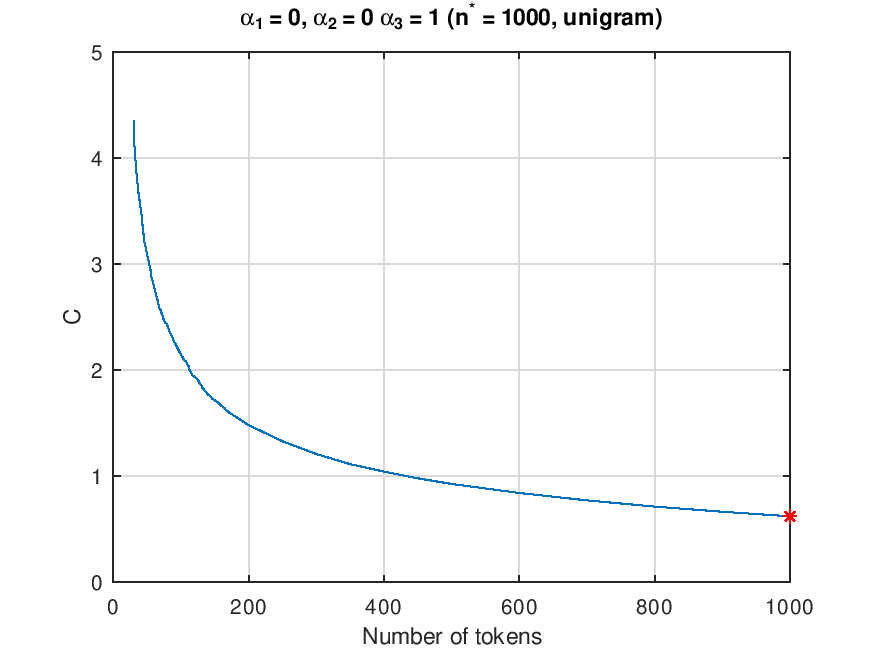}
    \caption{ $\alpha_{1,2,3} = (0,0,1)$; $n^*=1000$.}
    \label{fig:nstarsingle_unigram_c}
    \end{subfigure}
    \caption{\SP-\unigram. $n^*$ marked with a red "*".  x-axis shows the number of tokens and y-axis the $\C$.} 
    \label{fig:nstarsingle_unigram}
\end{figure*}

\begin{figure*}[!htb]
    \centering
    \begin{subfigure}{0.33\textwidth}
    \includegraphics[width=\textwidth]{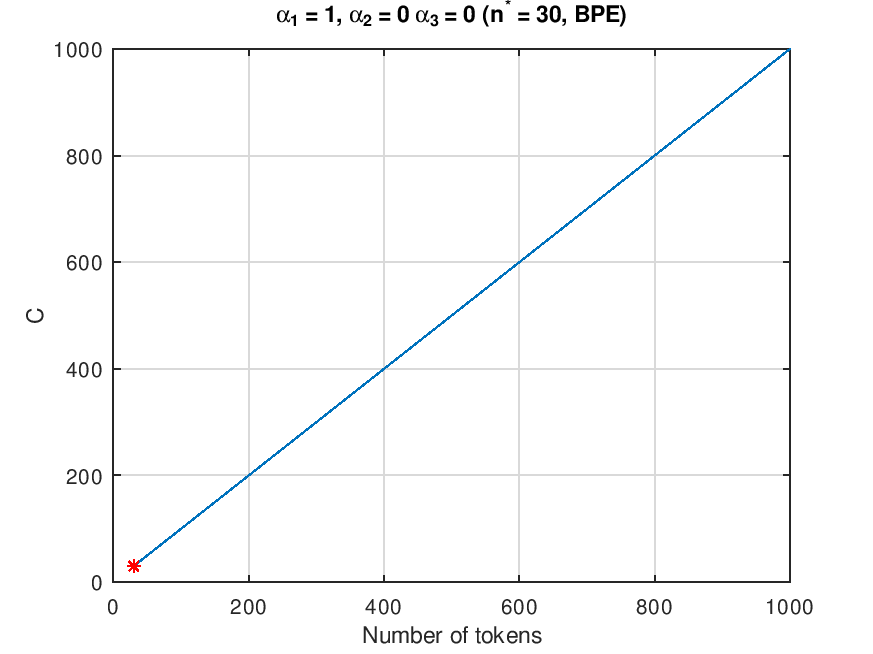}
    \caption{ $\alpha_{1,2,3} = (1,0,0)$; $n^*=30$.}
    \label{fig:nstarsingle_bpe_a}
    \end{subfigure}
    \hfill
    \begin{subfigure}{0.33\textwidth}
    \includegraphics[width=\textwidth]{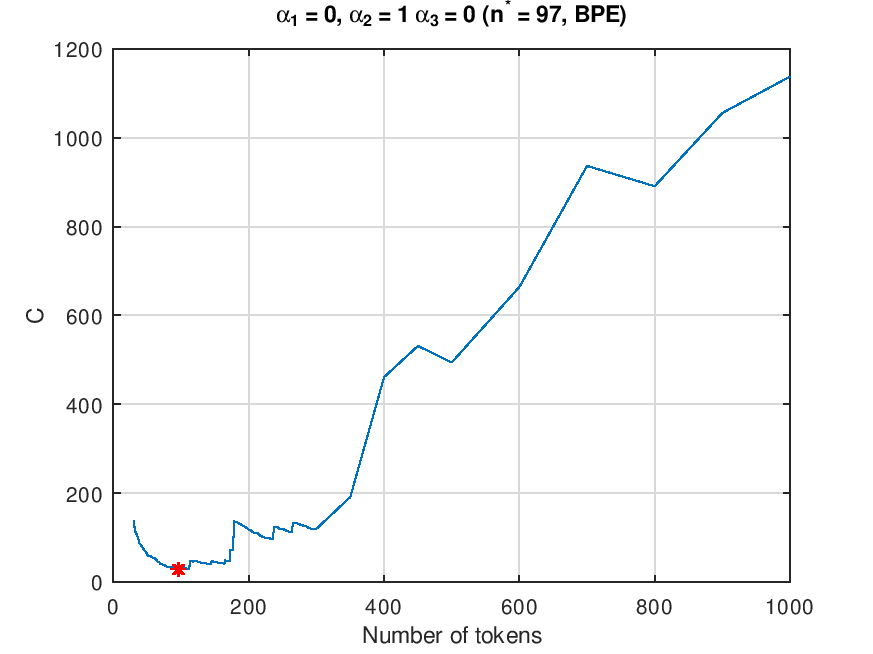}
    \caption{ $\alpha_{1,2,3} = (0,1,0)$; $n^*=97$.}
    \label{fig:nstarsingle_bpe_b}
    \end{subfigure}
    \hfill
    \begin{subfigure}{0.33\textwidth}
    \includegraphics[width=\textwidth]{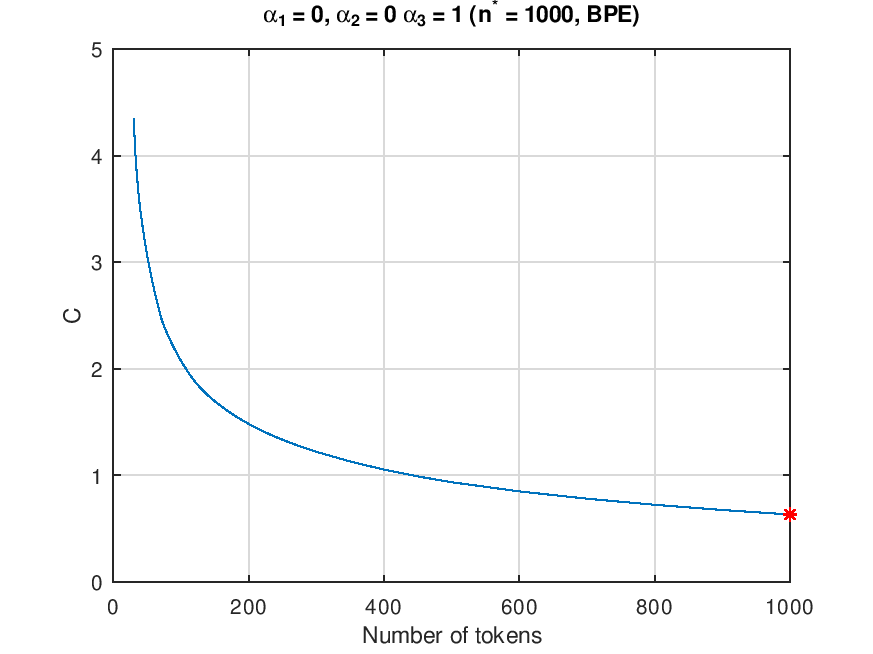}
    \caption{ $\alpha_{1,2,3} = (0,0,1)$; $n^*=1000$.}
    \label{fig:nstarsingle_bpe_c}
    \end{subfigure}
    \caption{\SP-\bpe. $n^*$ marked with a red "*". x-axis shows the number of tokens and y-axis the $\C$.} 
    \label{fig:nstarsingle_bpe}
\end{figure*}

\begin{figure*}[!htb]
    \centering
    \begin{subfigure}{0.475\textwidth}
    \includegraphics[width=\textwidth]{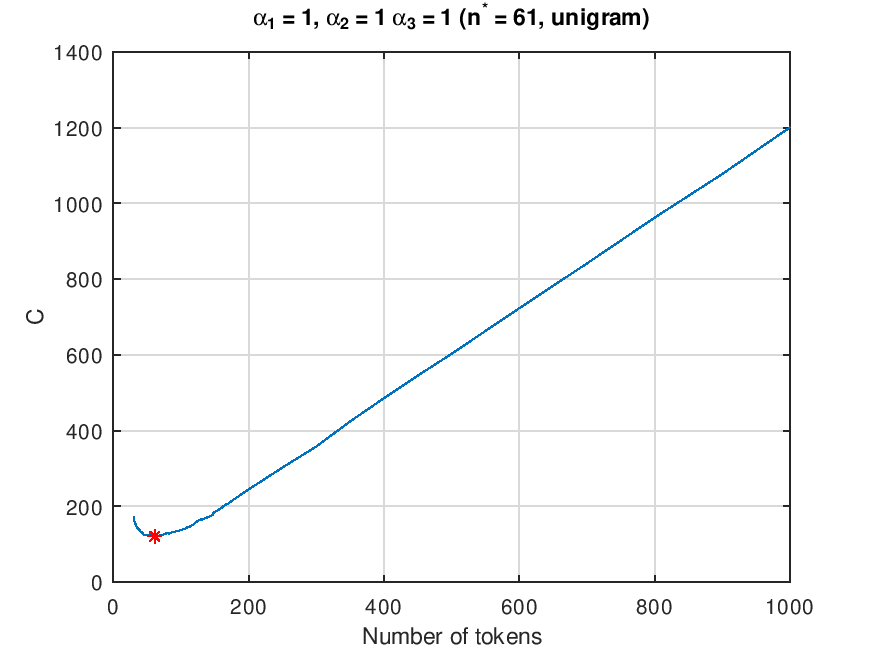}
    \caption{\SP-\unigram. $n^*=61$.}
    \label{fig:f_1_1_1_unigram}
    \end{subfigure}
    \hfill
    \begin{subfigure}{0.475\textwidth}
    \includegraphics[width=\textwidth]{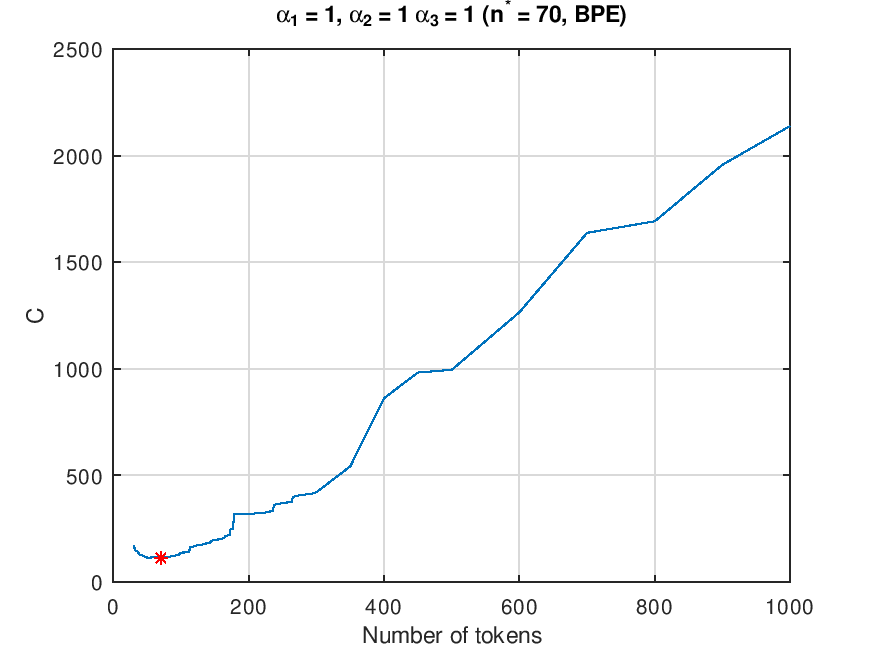}
    \caption{\SP-\bpe.  $n^*=70$.} 
    \label{fig:f1_1_1_bpe}
    \end{subfigure}
    \caption{Optimal number of tokens ($n^*$) for $\alpha_{1,2,3} = (1,1,1)$}
   \label{fig:f1_1_1}
\end{figure*}

    We computed the cost function $\C$ (\ref{eq:mimimize}) by varying $n$ between $30$ and $1000$. We found that higher value of $n (>1000)$ always resulted in a higher $\C$ and $n<30$ resulted in the number of tokens ($n$) being less than the number of characters in the training text data. 
    We trained the \SP\ \cite{kudo2018sentencepiece}, with unigram 
    (--vocab\_size=n, --model\_type=unigram, --split\_by\_whitespace=False) called \SP-\unigram\ model 
    and with BPE 
    (--vocab\_size=n, --model\_type=bpe, --split\_by\_whitespace=False) called \SP-\bpe\ model.
    %
    Figure \ref{fig:all_data} shows the plot of $t_1, t_2, t_3$ (\ref{eq:mimimize}) as a function of number of \sub-words\ (x-axis). 
    In both Figure \ref{fig:all_data_unigram} (\SP-\unigram) and Figure \ref{fig:all_data_bpe} (\SP-\bpe) it can be observed that 
    while $t_1$ is linearly increasing and $t_3$ is exponentially decreasing; $t_2$ exhibits 
    a dip before linearly increasing as a function of $n$.  Figure \ref{fig:nstarsingle_unigram} (Figure \ref{fig:nstarsingle_bpe}) shows the plot of the cost function $C$, 
    for \SP-\unigram\ (\SP-\bpe), for $n=1, \cdots, 1000$ for $\alpha_{1,2,3}=1,0,0$ (Figure \ref{fig:nstarsingle_unigram_a} (\ref{fig:nstarsingle_bpe_a})), $\alpha_{1,2,3}=0,1,0$ (Figure \ref{fig:nstarsingle_unigram_b} (\ref{fig:nstarsingle_bpe_b})), and $\alpha_{1,2,3}=0,0,1$ (Figure \ref{fig:nstarsingle_unigram_c} (\ref{fig:nstarsingle_bpe_c})) with the $n^*$ (\ref{eq:nstar}) marked with a red "*". Clearly the minimum
    value 
    of 
    $\C$ (\ref{eq:mimimize}) varies with different values of $\alpha$'s. As expected, $n^* =30$ when the cost function is only a function of the number of \sub-words\ for both \SP-\unigram\ (Figure \ref{fig:nstarsingle_unigram_a}) and \SP-\bpe\ (Figure \ref{fig:nstarsingle_bpe_a}) and  $n^* =1000$ when  $\C$ is a function of only $t_3$ (see Figure \ref{fig:nstarsingle_unigram_c} and \ref{fig:nstarsingle_bpe_c}). The most interesting aspect is observable for $t_2$ where the $n^*=145$ for \SP-\unigram\ (see Figure \ref{fig:nstarsingle_unigram_b}) and $n^*=97$ for \SP-\bpe\  (see Figure \ref{fig:nstarsingle_bpe_b}). Figure \ref{fig:f1_1_1} shows the plot of the cost function $\C$ for $\alpha_{1,2,3}=1$.  As shown in Figure \ref{fig:f_1_1_1_unigram} the minimum value of $\C$ occurs for $n^*=61$ for \SP-\unigram\ model while $n^* = 70$ for \SP-\bpe\ model. 

The experiments above provide the optimal values for number of \sub-words corresponding 
to different 
values of $\alpha$'s. We now present the performance of the ASR systems with the number of \sub-words obtained above, 
namely $n^* = 30, 61, 145, 1000$ for \SP-\unigram\ and 
$n^* = 30, 97, 70, 1000$ for \SP-\bpe,
in Table \ref{tab:results}. 
We compare this with 
the
ESPNet recipe 
recommendation of using 
the number of \sub-words as 300 
using \SP-\unigram\ 
language model. 
Using this value ($n=300$) results in an average WERs of $13.8\%$ over the dev sets and $14.5\%$ over the test sets (see Table \ref{tab:results}). Using 30 \sub-words (which corresponds to $\alpha_{1,2,3} = (1,0,0)$, the average WER over test sets reduces very slightly to $14.3\%$, however the average WER over the dev sets increases 
slightly to $13.9\%$. Using $n^*=61$ \sub-words and 1000 \sub-words (corresponding to $\alpha_{1,2,3}=(0,1,0)$ and $(0,0,1)$, respectively), did not improve the performance, with $n=1000$ \sub-words performing the worst. It suggests that blindly increasing the number of \sub-words does not 
necessarily
improve the performance of the ASR. When we provide equal weight of $1$ to $t_1, t_2$ and $t_3$ in the cost function, we get the number of \sub-words as $n^*=145$ (\SP-\unigram) and $n^*=70$ (\SP-\bpe). This 
choice of $n$ for \SP-\unigram\ 
outperforms all the other systems resulting in an average WER of $13.2\%$ over the dev sets and an average WER of $13.6\%$ over the test sets.
We observe similar improvement in ASR performance for \SP-\bpe\ . While the
recommended $n=300$ \sub-words with ASR result in an average WERs of $14.2\%$ and $14.4\%$ with dev and test sets respectively, the choice of $n^*=70 $ \sub-words, which correspond to $\alpha_{1,2,3} = (1,1,1)$ show an improved performance of $13.7\%$ and $14.2\%$ WER for dev and test sets respectively.
 What is remarkable is that this improvement in ASR
performance comes with reduced computational cost when compared to the 
use of 
recommended $300$ \sub-words
in ESPNet recipe. 
%
This set of experiments again reinforces the earlier finding that blindly increasing the number of \sub-words does not improve the performance of the ASR system.

\begin{table}[t]
\caption{WERs (in \%) for various $n$ on Librispeech 100h. ``dev-avg": average over dev-clean and dev-other. ``test-avg": average over test-clean and test-other}
\label{tab:results}
\begin{center}
\resizebox{\columnwidth}{!}{
\begin{tabular}{||c||cc|c||cc|c||} \hline
\multicolumn{7}{||c||}{\SP-\unigram} \\ \hline
$n (\alpha_{1,2,3})$ & dev-clean & dev-other & dev-avg & test-clean & test-other & test-avg \\ \hline
300 (-)& 7.7 & 20.0 & 13.8 & 8.3 & 20.8 & 14.5 \\ \hline
30 (1, 0, 0)   & 7.6  & 20.3 & 13.9 & 7.9 & 20.7 & 14.3\\ 
145 (0, 1, 0)& 8.6 & 20.8 & 14.7 & 9.0 & 21.0 & 15.0\\
1000 (0, 0, 1)& 11.0 & 21.0 & 16.0 & 11.5 & 23.0 & 17.2\\ 
{\bf 61 (1, 1, 1)}& {\bf 7.2} & {\bf 19.2} & {\bf 13.2} & {\bf 7.7} & {\bf 19.6} & {\bf 13.6}\\ 
\hline
\multicolumn{7}{||c||}{\SP-\bpe} \\ \hline
$n (\alpha_{1,2,3})$ & dev-clean & dev-other & dev-avg & test-clean & test-other & test-avg \\ \hline
300 (-)& 8.1 & 20.4 & 14.2 & 8.5 & 20.4 & 14.4 \\ \hline
30 (1, 0, 0) & 7.6 & 20.3 & 13.9 & {\bf 7.9} & 20.7 & 14.3\\ 
97 (0, 1, 0)& 7.7 & 20.1 & 13.9 & 8.2 & 20.4 & 14.3\\
1000 (0, 0, 1)& 7.9 & 20.0 & 13.9 & 8.0 & 20.6 & 14.3\\ 
{\bf 70 (1, 1, 1)}& {\bf 7.6} & {\bf 19.8} & {\bf 13.7 } & 8.0 & {\bf 20.4} & {\bf 14.2}\\ \hline
\end{tabular}
}
\end{center}
\end{table}

It is interesting to note that the best performance is obtained when $\alpha_{1,2,3} = 1$.
This suggests that each of the terms $t_1, t_2, t_3$ in the cost function has a role to play in determining the number of tokens optimal for the ASR performance. The number of tokens will change when the training text changes or the tokenizer changes and hence  the cost function minimization should be performed once every time the training text or the tokenizer changes. 

\section{Conclusions}
\label{sec:conclude}
In this paper we proposed a formulation based on construction of a cost function that allows for the identification of an optimal vocabulary size for the tokenizers used during the training of a speech recognition engine. The formulated cost function is based on keeping a balance between the most frequently and least frequently occurring training sub-word data to avoid bias in training as well as making sure that the number of tokens required to represent the data is minimal from the computing cost perspective. Using Librispeech 100 hours training set, we showed the efficacy of the approach to determining the vocabulary size of the tokenkinzer. In future, it would be worthwhile to look into the cost function in more detail to include other relevant factors. A robust approach to determining the values of $\alpha$'s would improve the performance further.

\bibliographystyle{IEEEtran}
\bibliography{mybib,refs.bib}

\begin{thebibliography}{10}
\providecommand{\url}[1]{#1}
\csname url@samestyle\endcsname
\providecommand{\newblock}{\relax}
\providecommand{\bibinfo}[2]{#2}
\providecommand{\BIBentrySTDinterwordspacing}{\spaceskip=0pt\relax}
\providecommand{\BIBentryALTinterwordstretchfactor}{4}
\providecommand{\BIBentryALTinterwordspacing}{\spaceskip=\fontdimen2\font plus
\BIBentryALTinterwordstretchfactor\fontdimen3\font minus
  \fontdimen4\font\relax}
\providecommand{\BIBforeignlanguage}[2]{{%
\expandafter\ifx\csname l@#1\endcsname\relax
\typeout{** WARNING: IEEEtran.bst: No hyphenation pattern has been}%
\typeout{** loaded for the language `#1'. Using the pattern for}%
\typeout{** the default language instead.}%
\else
\language=\csname l@#1\endcsname
\fi
#2}}
\providecommand{\BIBdecl}{\relax}
\BIBdecl

\bibitem{raissi2021consistent}
T.~Raissi, E.~Beck, R.~Schlüter, and H.~Ney, ``Towards consistent hybrid hmm
  acoustic modeling,'' arXiv, 2021.

\bibitem{sennrich-bpe}
\BIBentryALTinterwordspacing
R.~Sennrich, B.~Haddow, and A.~Birch, ``Neural machine translation of rare
  words with subword units,'' in \emph{Proceedings of the 54th Annual Meeting
  of the Association for Computational Linguistics (Volume 1: Long Papers)},
  K.~Erk and N.~A. Smith, Eds.\hskip 1em plus 0.5em minus 0.4em\relax Berlin,
  Germany: Association for Computational Linguistics, Aug. 2016, pp.
  1715--1725. [Online]. Available: \url{https://aclanthology.org/P16-1162}
\BIBentrySTDinterwordspacing

\bibitem{wordpiece}
M.~Schuster and K.~Nakajima, ``Japanese and korean voice search,'' in
  \emph{2012 IEEE International Conference on Acoustics, Speech and Signal
  Processing (ICASSP)}, 2012, pp. 5149--5152.

\bibitem{unigram}
\BIBentryALTinterwordspacing
T.~Kudo, ``Subword regularization: Improving neural network translation models
  with multiple subword candidates,'' in \emph{Proceedings of the 56th Annual
  Meeting of the Association for Computational Linguistics (Volume 1: Long
  Papers)}, I.~Gurevych and Y.~Miyao, Eds.\hskip 1em plus 0.5em minus
  0.4em\relax Melbourne, Australia: Association for Computational Linguistics,
  Jul. 2018, pp. 66--75. [Online]. Available:
  \url{https://aclanthology.org/P18-1007}
\BIBentrySTDinterwordspacing

\bibitem{kudo2018sentencepiece}
\BIBentryALTinterwordspacing
T.~Kudo and J.~Richardson, ``{S}entence{P}iece: A simple and language
  independent subword tokenizer and detokenizer for neural text processing,''
  Brussels, Belgium, pp. 66--71, Nov. 2018. [Online]. Available:
  \url{https://aclanthology.org/D18-2012}
\BIBentrySTDinterwordspacing

\bibitem{pasm}
H.~Xu, S.~Ding, and S.~Watanabe, ``Improving end-to-end speech recognition with
  pronunciation-assisted sub-word modeling,'' in \emph{ICASSP 2019 - 2019 IEEE
  International Conference on Acoustics, Speech and Signal Processing
  (ICASSP)}, 2019, pp. 7110--7114.

\bibitem{pis}
\BIBentryALTinterwordspacing
V.~Papadourakis, M.~Mueller, J.~Liu, A.~Mouchtaris, and M.~Omologo,
  ``Phonetically induced subwords for end-to-end speech recognition,'' in
  \emph{Interspeech 2021}, 2021. [Online]. Available:
  \url{https://www.amazon.science/publications/phonetically-induced-subwords-for-end-to-end-speech-recognition}
\BIBentrySTDinterwordspacing

\bibitem{zouhar-etal-2023-formal}
\BIBentryALTinterwordspacing
V.~Zouhar, C.~Meister, J.~Gastaldi, L.~Du, T.~Vieira, M.~Sachan, and
  R.~Cotterell, ``A formal perspective on byte-pair encoding,'' in
  \emph{Findings of the Association for Computational Linguistics: ACL 2023},
  A.~Rogers, J.~Boyd-Graber, and N.~Okazaki, Eds.\hskip 1em plus 0.5em minus
  0.4em\relax Toronto, Canada: Association for Computational Linguistics, Jul.
  2023, pp. 598--614. [Online]. Available:
  \url{https://aclanthology.org/2023.findings-acl.38}
\BIBentrySTDinterwordspacing

\bibitem{katare2024analyzing}
D.~Katare, D.~S. Noguero, S.~Park, N.~Kourtellis, M.~Janssen, and A.~Y. Ding,
  ``Analyzing and mitigating bias for vulnerable classes: Towards balanced
  representation in dataset,'' arXiv, 2024.

\bibitem{watanabe18_interspeech}
S.~Watanabe, T.~Hori, S.~Karita, T.~Hayashi, J.~Nishitoba, Y.~Unno, N.~{Enrique
  Yalta Soplin}, J.~Heymann, M.~Wiesner, N.~Chen, A.~Renduchintala, and
  T.~Ochiai, ``{ESPnet: End-to-End Speech Processing Toolkit},'' in \emph{Proc.
  Interspeech 2018}, 2018, pp. 2207--2211.

\bibitem{kingma2017adam}
\BIBentryALTinterwordspacing
D.~P. Kingma and J.~Ba, ``Adam: {A} method for stochastic optimization,'' in
  \emph{3rd International Conference on Learning Representations, {ICLR} 2015,
  San Diego, CA, USA, May 7-9, 2015, Conference Track Proceedings}, Y.~Bengio
  and Y.~LeCun, Eds., 2015. [Online]. Available:
  \url{http://arxiv.org/abs/1412.6980}
\BIBentrySTDinterwordspacing

\bibitem{vas_attn}
\BIBentryALTinterwordspacing
A.~Vaswani, N.~Shazeer, N.~Parmar, J.~Uszkoreit, L.~Jones, A.~N. Gomez, L.~u.
  Kaiser, and I.~Polosukhin, ``Attention is all you need,'' in \emph{Advances
  in Neural Information Processing Systems}, I.~Guyon, U.~V. Luxburg,
  S.~Bengio, H.~Wallach, R.~Fergus, S.~Vishwanathan, and R.~Garnett, Eds.,
  vol.~30.\hskip 1em plus 0.5em minus 0.4em\relax Curran Associates, Inc.,
  2017. [Online]. Available:
  \url{https://proceedings.neurips.cc/paper_files/paper/2017/file/3f5ee243547dee91fbd053c1c4a845aa-Paper.pdf}
\BIBentrySTDinterwordspacing

\bibitem{ko15_interspeech}
T.~Ko, V.~Peddinti, D.~Povey, and S.~Khudanpur, ``{Audio augmentation for
  speech recognition},'' in \emph{Proc. Interspeech 2015}, 2015, pp.
  3586--3589.

\bibitem{park19e_interspeech}
D.~S. Park, W.~Chan, Y.~Zhang, C.-C. Chiu, B.~Zoph, E.~D. Cubuk, and Q.~V. Le,
  ``{SpecAugment: A Simple Data Augmentation Method for Automatic Speech
  Recognition},'' in \emph{Proc. Interspeech 2019}, 2019, pp. 2613--2617.

\end{thebibliography}

\end{document}